\documentclass[twocolumn,pra,superscriptaddress,longbibliography]{revtex4-1}
\usepackage{graphicx}
\usepackage{amssymb}
\usepackage{amsmath}
\usepackage{color}
\usepackage{comment}

\newcommand{\vect}[1]{{\mathbf #1}}
\renewcommand{\k}{{\bf k}}

\newcommand{\am}{a_-}

\newcommand{\ek}{\epsilon_{\k}}

\newcommand{\nn}{\nonumber}

\newcommand{\F}{{\cal F}}
\newcommand{\cz}{{\cal C}_z}
\newcommand{\T}{{\cal T}}
\newcommand{\op}{\omega_z}

\def\lba{\left(}    \def\rba{\right)}
\def\lbc{\left[}    \def\rbc{\right]}

\newcommand{\sout}[1]{}

\begin{document}

%\linenumbers

\title{Efimov trimers under strong confinement}

\author{Jesper Levinsen}
\affiliation{Aarhus Institute of Advanced Studies, Aarhus University, DK-8000 Aarhus C, Denmark}
\affiliation{Cavendish Laboratory, JJ Thomson Avenue, Cambridge,
 CB3 0HE, United Kingdom}

\author{Pietro Massignan}
\affiliation{ICFO - The Institute of Photonic Sciences - 08860 Castelldefels (Barcelona), Spain}

\author{Meera M. Parish}
\affiliation{London Centre for Nanotechnology, Gordon Street, London, WC1H 0AH, United Kingdom}

\date{\today}

\begin{abstract}
  The dimensionality of a system can fundamentally impact the
  behaviour of interacting quantum particles. Classic examples range
  from the fractional quantum Hall effect to high temperature
  superconductivity. As a general rule, one expects confinement to
  favour the binding of particles.  However, attractively interacting
  bosons apparently defy this expectation: while three identical
  bosons in three dimensions can support an infinite tower of Efimov
  trimers, only two universal trimers exist in the two dimensional
  case. Here we reveal how these two limits are connected by
  investigating the problem of three identical bosons confined by a
  harmonic potential along one direction.  We show that the
  confinement breaks the discrete Efimov scaling symmetry and
  successively destroys the weakest bound trimers. However, the
  deepest bound trimers persist even under strong confinement. In
  particular the ground state Efimov trimer hybridizes with the
  two-dimensional trimers, yielding a superposition of trimer
  configurations that effectively involves tunnelling through a
  short-range repulsive barrier.  Our results suggest a way to use
  strong confinement to engineer more stable Efimov-like trimers,
  which have so far proved elusive.
\end{abstract}

%\begin{abstract}
%  The dimensionality of a system can fundamentally impact the
%  behaviour of interacting quantum particles. Classic examples range
%  from the fractional quantum Hall effect to high temperature
%  superconductivity. As a general rule, one expects confinement to
%  favour the binding of particles.  However, attractively interacting
%  bosons apparently defy this expectation: while three identical
%  bosons in three dimensions can support an infinite tower of Efimov
%  trimers, only two universal trimers exist in the two dimensional
%  case. Here we reveal how these two limits are connected by
%  investigating the problem of three identical bosons confined by a
%  harmonic potential along one direction. We show that the confinement
%  breaks the discrete Efimov scaling symmetry and destroys the weakest
%  bound trimers.  However, the deepest bound Efimov trimer persists
%  under strong confinement and hybridizes with the
%  quasi-two-dimensional trimers, yielding a superposition of trimer
%  configurations that effectively involves tunnelling through a
%  short-range repulsive barrier.  Our results suggest a way to use
%  strong confinement to engineer more stable Efimov-like trimers,
%  which have so far proved elusive.
%\end{abstract}

\maketitle

\section{Introduction}
The quantum mechanical three-body problem displays surprisingly rich
and complex behaviour despite its apparent simplicity. A fundamental
example is the Efimov effect~\cite{Efimov1970}, which has intrigued
physicists for decades owing to its peculiar and universal scaling
properties. Its simplest incarnation only requires three bosons with
resonant short-range interactions, and it can thus occur in a wide
variety of systems ranging from nucleon systems~\cite{Efimov1971} and
ultracold atomic gases~\cite{Ferlaino2011}, to quantum
magnets~\cite{Nishida2013}.  In particular, the cold-atom system
finally provided the first experimental evidence for Efimov physics in
2006~\cite{Kraemer2006}, thus stimulating even greater interest in the
subject.

A hallmark of the Efimov effect is a spectrum of three-body bound
states (trimers) that exhibits a {\em discrete} scaling symmetry: for the
simple case of three-dimensional (3D) identical bosons, the energy $E$
of one trimer can be mapped onto another via the transformations $E
\to \lambda_0^{-2n} E$ and $a \to \lambda_0^n a$, where $a$ is the
two-body scattering length, $\lambda_0$ is a known factor, and $n$ is
any integer~\cite{Braaten2006}. In particular, at the unitarity point
$1/a = 0$, the scattering length drops out of the problem and there
exists an \emph{infinite} number of weakly-bound $s$-wave trimer
states~\cite{Efimov1970}, with the deepest bound trimer set by the
short distance physics~\cite{Braaten2006}.  Such self-similar
behaviour is reminiscent of more complex systems in Nature, such as
coastlines, snow flakes and ferns~\cite{mandelbrot1983fractal}, rather
than of a typical few-body system --- for instance, the two-body
problem only exhibits a \emph{continuous} scaling symmetry, where the
low-energy properties simply scale with $a$.  It is then natural to
ask how these Efimov trimers evolve once the bosons are subject to
confinement and the motion is constrained.

\begin{center}
\begin{figure}[ht]
\vskip 0 pt
\includegraphics[width=\columnwidth,clip]{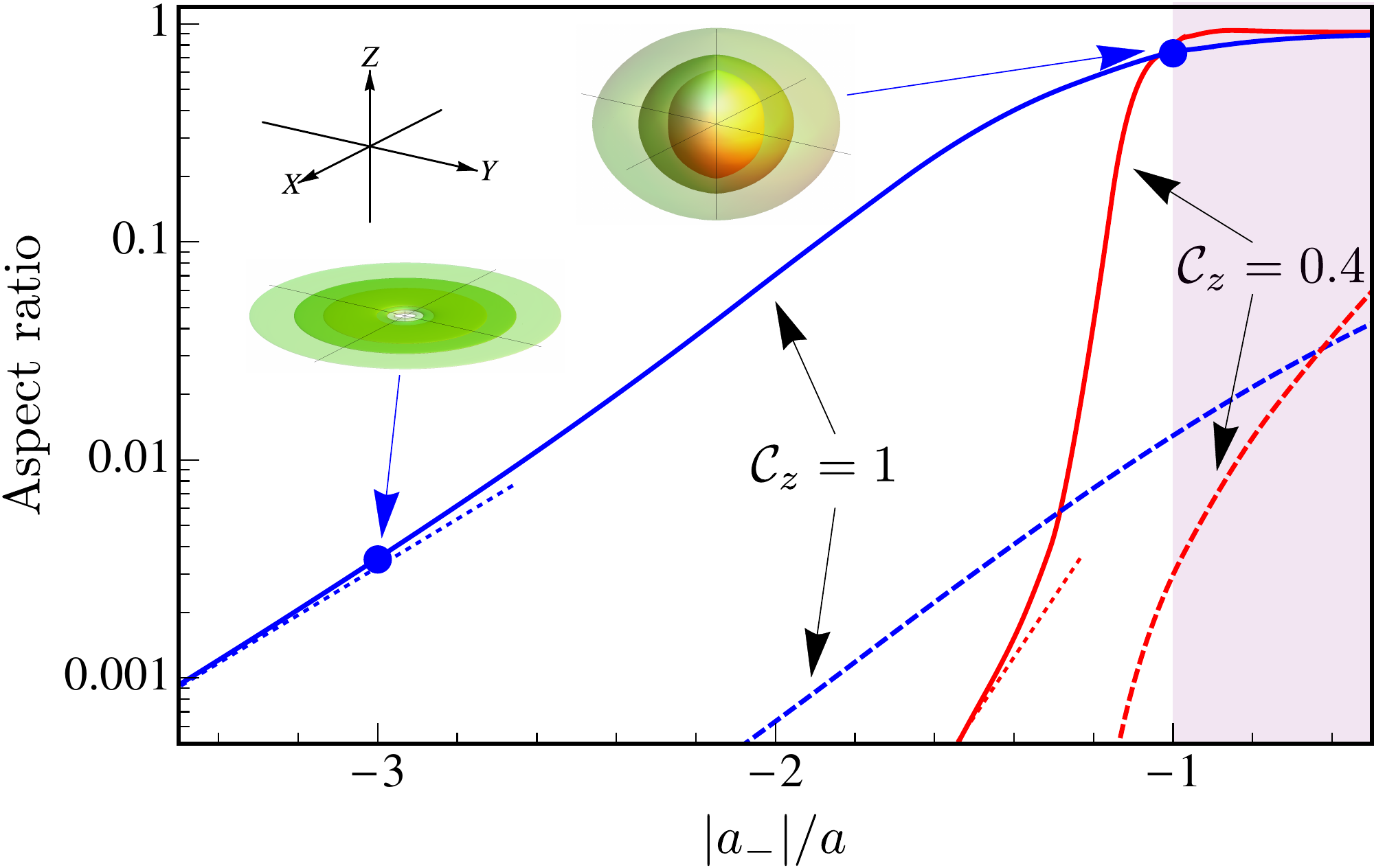}
\caption{Shape of the trimers under confinement: The aspect ratio
  $2\langle Z^2\rangle/\langle\rho^2\rangle$ for the two deepest
  trimers is shown as a function of interaction for two different
  confinement strengths ${\cal C}_z$. Here $\boldsymbol{\rho}$ is the
  separation of an atom and a pair in the $x$-$y$ plane, and $Z$ is
  the separation in the confined direction (see main text).  The
  deepest trimer (solid lines) only exists in 3D for $|\am|/a\geq-1$,
  as depicted by the shaded region.  Here the aspect ratio is close to
  1, indicating that the trimer wavefunction resembles the 3D Efimov
  state. Outside the trimer's regime of existence in 3D, the aspect
  ratio quickly decreases, indicating that the trimer spreads out in
  the 2D plane.  The change in aspect ratio as $a$ crosses $\am$ is
  more gradual for the stronger confinement, due to a stronger
  coupling between trimers of a 2D and 3D character.  The deepest
  trimer eventually approaches the 2D asymptotic limit (dotted lines,
  see Appendix \ref{app:AR}).  A similar picture emerges for the first
  excited trimer (dashed lines), but the aspect ratio here is much
  smaller than 1 even away from the 2D limit.  The shape of the
  deepest trimer is illustrated in the insets, which show surface
  density plots of the squared wavefunction (see the main text)
  evaluated at the points marked by filled circles.}
\label{fig:aspect}
\end{figure}
\end{center}

Cold-atom experiments already require the presence of a weak trapping
potential, but the remarkable tunability of the atomic system has
meant that more extreme versions of confinement can now be realised,
where one can create 2D Bose gases with markedly different many-body
properties \cite{Hadzibabic2006,Clade2009,Hung2011}.  It is already
known that the system dimensionality radically changes the few-boson
problem: in 2D, the Efimov effect is absent~\cite{Nishida2011} and
only two $s$-wave trimers are predicted to exist, with universal
energies $-16.5 |E_b|$ and $-1.27 |E_b|$~\cite{Bruch1979}, where $E_b$
is the two-body (dimer) binding energy.  Here we show how this 2D
limit evolves into Efimovian behaviour as we relax the confinement.
We consider the simplest scenario of three identical bosons of mass
$m$ subjected to a tight harmonic confinement in the $z$-direction,
$V(z)=\frac{1}{2}m\omega_z^{2}z^{2}$, with confinement frequency
$\omega_z$.  While the weakest bound Efimov trimers are successively
destroyed with increasing confinement, crucially we find that the
deepest states persist even for strong confinement. In particular, the
ground state trimer is actually stabilised beyond its original regime
of existence in 3D (Fig.~\ref{fig:aspect}).  Moreover, in contrast to
the dimer case, we obtain avoided crossings in the trimer spectrum. By
evaluating the three-body hyperspherical potentials, we show that the
avoided crossings correspond to trimer states that are superpositions
of both short-range 3D-like and long-range 2D-like trimer
configurations separated by a repulsive barrier.  Such hybrid trimers
could potentially be used to manufacture more stable Efimov-like
trimers, thus paving the way for the exploration of many-body states
of trimers.

%While the confinement in general destroys the weakest
%bound Efimov trimers, crucially we find that the deepest state
%persists even for strong confinement, and is actually stabilised
%beyond its original regime of existence by acquiring a long-range 2D
%component (Fig.~\ref{fig:aspect}).  Moreover, in contrast to the dimer
%case, we obtain avoided crossings in the trimer spectrum. These
%correspond to trimer states that are superpositions of both
%short-range 3D and long-range 2D trimer configurations separated by a
%repulsive barrier.  Such hybrid trimers could potentially be used to
%manufacture more stable Efimov-like trimers, thus paving the way for
%the exploration of many-body states of trimers.

\section{The quasi-2D three-boson problem}
The strong harmonic confinement described above is readily achieved in
the cold-atom system via the application of an optical lattice or
anisotropic trap~\cite{Hadzibabic2006,Clade2009,Hung2011}.  Indeed,
2D-3D crossovers have already been investigated in this manner in
Fermi gases~\cite{Sommer2012,Guenter2005}. For temperatures $T\ll
\omega_z$ (we set $\hbar=k_B=1$), non-interacting bosons will occupy
the lowest harmonic oscillator level and will thus be kinematically
2D. However, in the presence of boson-boson interactions, the
particles may virtually explore all excited states of the harmonic
potential; thus we refer to the confined system as
\emph{quasi}-two-dimensional (q2D).  An advantage of the harmonic
potential is that one can decouple the centre-of-mass motion from the
relative motion of the particles, so in the following we ignore the
centre-of-mass contribution.

The effect of q2D confinement is twofold: it introduces an extra
length scale, $l_z=\sqrt{1/m\omega_z}$, and it raises the threshold of
the three-atom continuum from 0 to $\omega_z$.  Assuming that $l_z$
and the scattering length $a$ greatly exceed the van der Waals range
of the interaction, the two-body problem is then completely
parametrised by the dimensionless quantity $l_z/a$, and there is
always a dimer bound state, in contrast to the 3D case.  For weak
attraction $l_z/a\ll -1$, we recover the 2D limit with dimer binding
energy $E_b = -\frac{B}{m \pi l_z^2} e^{\sqrt{2\pi}l_z/a}$, where
$B\approx0.905$~\cite{Petrov2001}, while for strong attraction
$l_z/a\gg1$ (or weak confinement), this evolves into the 3D binding
energy, $-1/ma^2$ (see Appendix \ref{app:H}).  The latter
corresponds to the regime where the dimer is much smaller than the
confinement length $l_z$ and is therefore barely perturbed by the
confinement.

The three-body problem, however, requires the additional length scale
$1/\kappa_*$, which is set by the short-distance physics and fixes the
3D trimer energies in the resonant limit: $E_T^{(n)}\approx
-\lambda_0^{-2n}\kappa^2_*/m$, with $n$ a positive integer and
$\lambda_0 \simeq 22.7$~\cite{Efimov1970}.  A more natural quantity to
consider in the cold-atom context is the scattering length $\am<0$ at
which the deepest Efimov trimer crosses the three-atom continuum: this
crossing leads to an enhanced three-body loss rate in the Bose gas,
which is the main observable in
experiment~\cite{Kraemer2006,zaccanti2009,Gross2009,Pollack2009}.
Moreover, there is a remarkable universal relationship between $\am$
and the van der Waals range~\cite{Berninger2011,Roy2013,Wang2012}.
Thus, we characterise the three-body problem using the interaction
parameter $|\am|/a$ and the confinement parameter ${\cal C}_z \equiv
|\am|/l_z$. Together, these determine how 2D or 3D a trimer is, as
encoded in the aspect ratio displayed in Fig.~\ref{fig:aspect}. In
particular, we see that when $|\am|/a < -1$ ({\em i.e.} when there are
no Efimov states in 3D), the two deepest trimers still persist under
confinement and become substantially flattened within the $x$-$y$
plane. Note that we only consider confinements $\cz \leq 1$, since
$\cz \gg 1$ will make our results nonuniversal and sensitive to the
details of the short-range interactions.

%In
%particular, we see that when $|\am|/a < -1$ and there are no Efimov
%states in 3D, the two deepest trimers still persist under confinement
%and become substantially flattened within the $x$-$y$ plane. Note that
%we only consider confinements $\cz \leq 1$, since $\cz \gg 1$ will
%make our results nonuniversal and sensitive to the details of the
%short-range interactions.

\begin{center}
\begin{figure*}[ht]
\vskip 0 pt
\includegraphics[width=0.95 \columnwidth, clip]{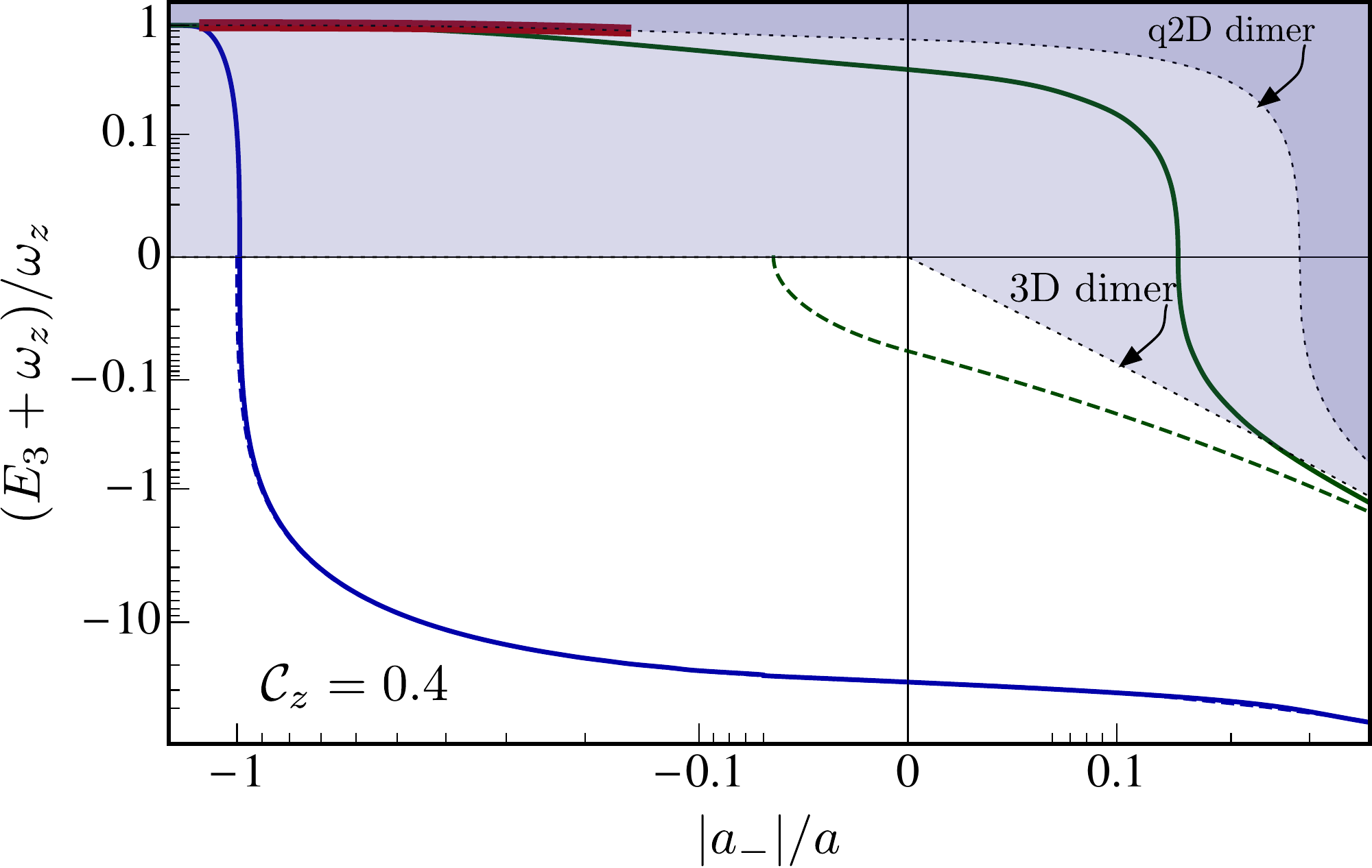}
\hskip 10 pt
\includegraphics[width=0.95 \columnwidth, clip]{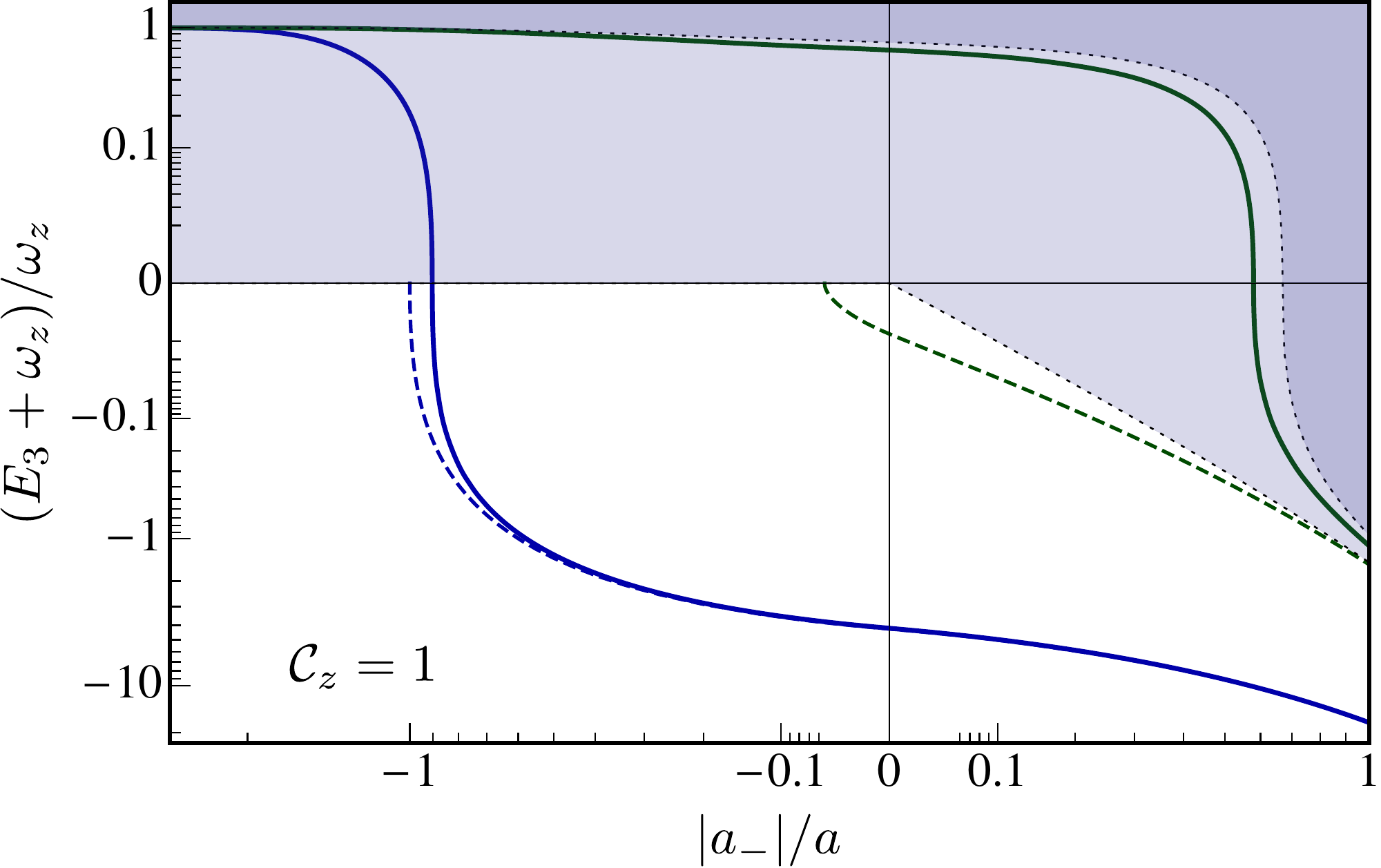}
\caption{Spectrum of trimers for two different confinement
    strengths. The q2D trimers are shown as solid lines, with
  energies $E_3 +\omega_z$ that take account of the raised three-atom
  continuum under confinement. For comparison, we also include the two
  deepest trimers in 3D (dashed lines). The shading illustrates the
  atom-dimer and three-atom continua, where the q2D continuum
  (contained within the 3D continuum) is shown with a darker
  shading. For a strong confinement with $\cz=1$ (right), only two q2D
  trimers exist. For a weaker confinement with $\cz=0.4$ (left), a
  third weakly bound trimer appears that is indistinguishable from the
  atom-dimer threshold on this scale; here we show its region of
  existence as a thick, red line on top of the threshold, but see
  Fig.~\ref{fig:2dstyle} for a clearer rendering.  In order to aid the
  visibility over the large energy range here, we rescaled the axes by
  means of the function $F_j(x)=\mbox{sgn}(x)\ln[1+5|x|^j]/\ln6$, with
  $j=1$ for the $x$-axis and $j=1/2$ for the $y$-axis. For the 3D
  spectrum, $\op$ is not defined and we display $m\am^2\cz^{-2} E_3$
  as a function of $|\am|/a$.}
\label{fig:3dstyle}
\end{figure*}
\end{center}

To determine the trimer wave functions and energies, we use the
Skorniakov--Ter-Martirosian (STM) equation, first introduced in the
context of neutron-deuteron scattering~\cite{stm}.  This takes
advantage of the short-range nature of the two-body interaction to
describe the three-body problem in terms of the relative motion of an
atom and a pair.  This equation has previously been extended to a q2D
geometry for two species of
fermions~\cite{Levinsen2009,Levinsen2013}. In the case of three
identical bosons, the STM equation for the q2D atom-pair vertex $\chi$
becomes
\begin{align} 
  &\T^{-1}\left(\k_1,E_3-\epsilon_{\k_1}
    -N_1\op\right)\chi_{\k_1}^{N_1 } \nn \\
  &=2\hspace{-3mm}\sum_{\k_2,N_2,n_{23},n_{31}} \hspace{-2mm}
  \frac{f_{n_{23}}f_{n_{31}} \langle N_1n_{23}|N_2n_{31}\rangle
    e^{-(k_1^2+k_2^2)/\Lambda^2}\chi_{\k_2}^{N_2}}
  {E_3-\epsilon_{\k_1}-\epsilon_{\k_2}-\epsilon_{\k_1+\k_2}-
    (N_1+n_{23})\op}.
\label{eq:stmq2d}
\end{align}
Here, the T-matrix $\T$ describes the repeated interaction of two
atoms, $E_3$ is the energy measured from the three-atom continuum
threshold, and $\ek=k^2/2m$. $\k_i$ is the relative momentum of atom
$i$ with the pair $(j,k)$ and we consider cyclic permutations of
$(i,j,k)=(1,2,3)$. Defining the relative motion in the $z$-direction
of two atoms, $z_{ij}=z_i-z_j$, and of an atom and a pair,
$z_{i,jk}=\frac{z_j+z_k}2-z_i$, the corresponding harmonic oscillator
quantum numbers are $n_{ij}$ and $N_i$. Then $\langle
N_1n_{23}|N_2n_{31}\rangle$ is the atom-pair Clebsch-Gordan
coefficient, with selection rule $N_1+n_{23}=N_2+n_{31}$, and
$f_{n_{ij}}$ is the relative harmonic oscillator wavefunction at
$z_{ij}=0$.  We include the short-distance physics by considering a
two-body separable potential of the form $e^{-(k^2+k'^2)/\Lambda^2}$,
where $\Lambda$ is an ultraviolet cutoff which fixes $\am$ (see
Appendix \ref{app:3bd}). Our results are independent of the specific
choice of cutoff as long as the relevant length scales, $|a|$ and
$l_z$, greatly exceed the short distance length scale $1/\Lambda$.
This is the case for all results presented in this Article.

The q2D three-boson problem presents a considerable challenge, owing
to the range of energy scales involved in the evolution towards
Efimovian behaviour.  
Since the 3D spectrum possesses a discrete energy scaling of
$22.7^{2}\approx 515$, in practice, we require at least $515^3$
Clebsch-Gordan coefficients after imposing the selection rule. 
However, the determination of these
coefficients is greatly simplified once one realises that they can be
related to Wigner's $d$-matrix~\cite{Wigner} as follows:
\begin{align} \label{eq:wigner}
\langle N_1n_{23}|N_2n_{31}\rangle =
  d^{\left(\frac{N_1+n_{23}}2\right)}_{\frac{N_2
      -n_{31}}2,\frac{N_1-n_{23}}2}(2\pi/3).
\end{align}
To see this, first note that $\langle N_1n_{23}|N_2n_{31}\rangle$ is
also the matrix element for the eigenstates of two isotropic 2D
harmonic oscillators, related by a rotation in the plane by $\pi/3$.
Then, using Schwinger's mapping~\cite{SchwingerQM}, one defines
angular momentum operators $\hat{\vect J}=\frac1%\hbar
2 \left(\begin{array}{cc} \hat{b}_1^\dag &
    \hat{b}_2^\dag\end{array}\right)
\hat{\boldsymbol\sigma} \left(\begin{array}{c}\hat{b}_1\\
    \hat{b}_2\end{array}\right)$, with $\hat{\boldsymbol\sigma}$ the
usual Pauli spin matrices and $\hat{b}_1$, $\hat{b}_2$ harmonic
oscillator operators. The eigenstates $\Theta$ of angular momentum are
$|\Theta(j,m)\rangle=|N_1 n_{23}\rangle$, where $j = \frac{N_1 +
  n_{23}}{2}$ and $m = \frac{N_1 -n_{23}}{2}$ are the usual quantum
numbers related to operators $\hat{J}^2$ and $\hat{J}_z$,
respectively.  In this basis, the rotation corresponds exactly to the
application of $e^{-i(2\pi/3) \hat{J}_y}$, and thus we obtain
Eq.~\eqref{eq:wigner}.

\section{Trimer spectra}

\begin{center}
\begin{figure*}[ht]
\vskip 0 pt
\includegraphics[width=0.95 \columnwidth, clip]{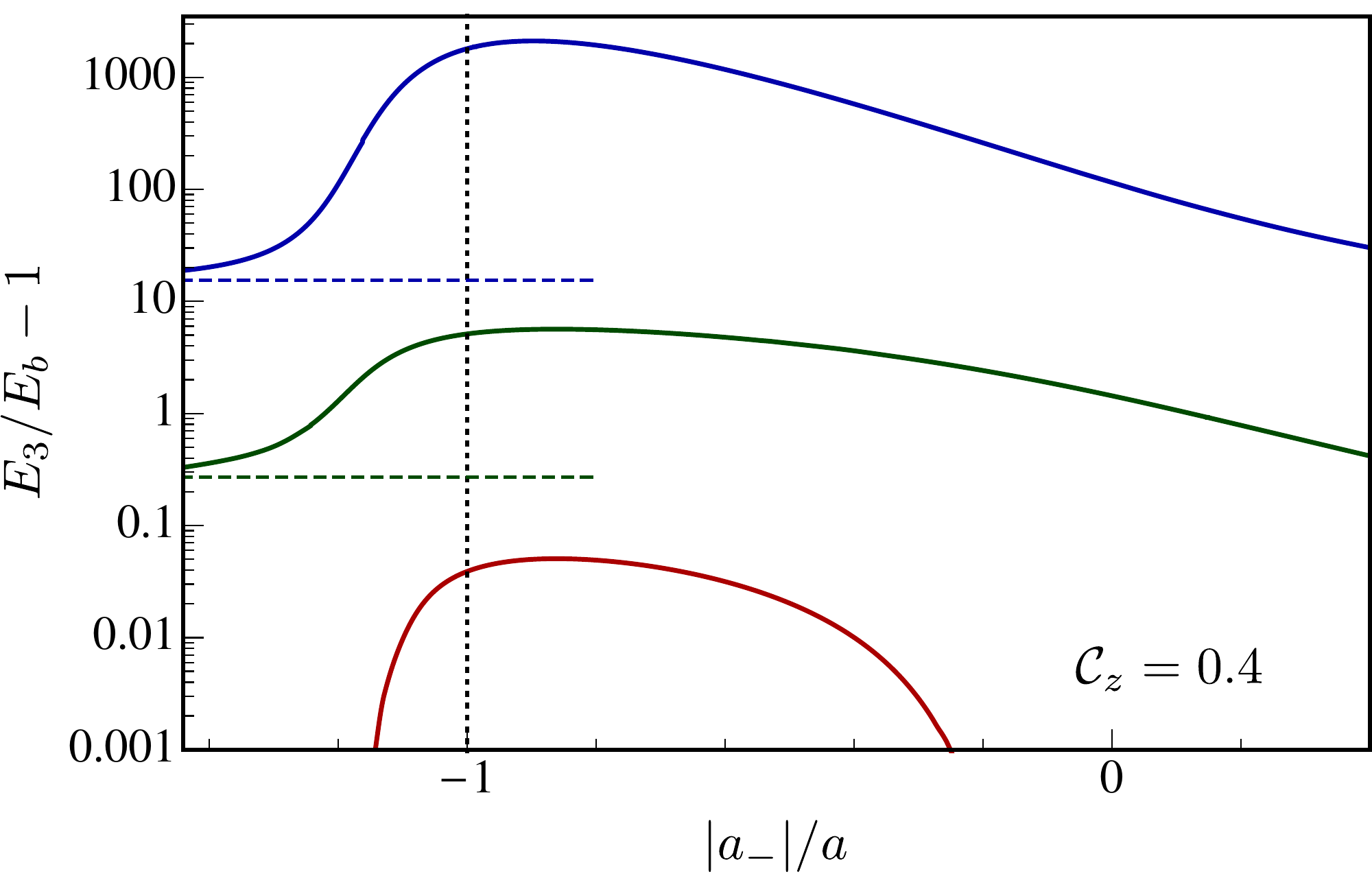}
\hskip 10 pt
\includegraphics[width=0.95 \columnwidth, clip]{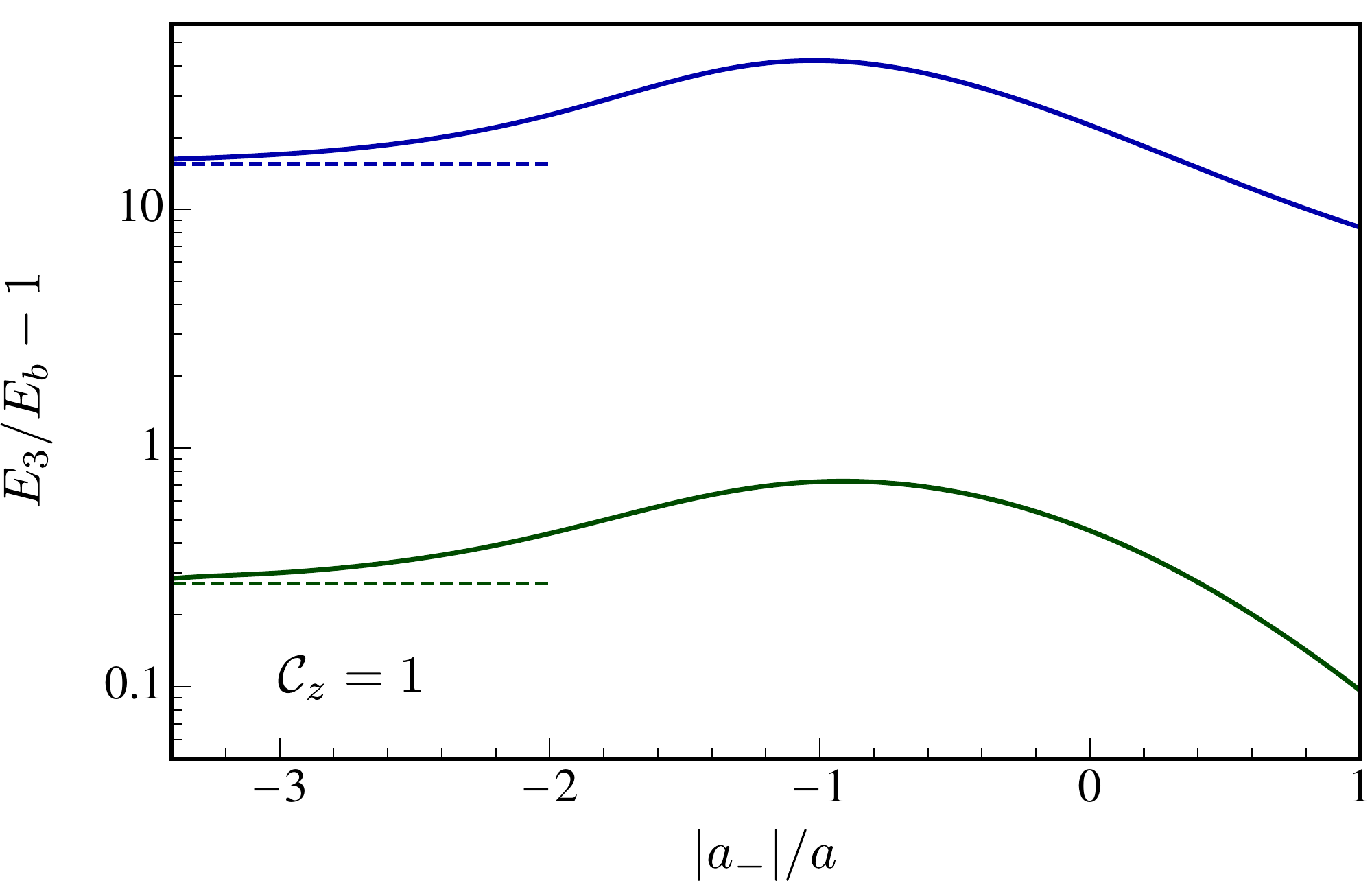}
\caption{Ratio between q2D trimer and dimer energies. The trimer
  energies (solid lines) are displayed as a function of interaction
  for two different confinements: $\cz=0.4$ (left) and $\cz=1$
  (right).  In the limit $|\am|/a\lesssim-1$, the trimer energies
  converge to the universal 2D results $-16.5|E_b|$ and
  $-1.27|E_b|$~\cite{Bruch1979} (dashed lines).  For moderate
  confinement (left), we see evidence of an avoided crossing resulting
  from the coupling between three `bare' states: the two 2D trimers,
  and a third trimer which emerges from the continuum at $a \simeq
  \am$ (vertical dotted line).}
\label{fig:2dstyle}
\end{figure*}
\end{center}

The trimer energies $E_3$ are found as non-trivial solutions of
Eq.~(\ref{eq:stmq2d}), and the complete spectra are displayed in
Figs.~\ref{fig:3dstyle} and \ref{fig:2dstyle}.  We see immediately
that the extra length scale $l_z$ removes the weakest bound Efimov
states, similarly to the effect of a finite scattering length in 3D.
In particular, $l_z$ may be interpreted as a large distance cut-off,
since Efimov trimers much larger than this will be strongly perturbed
by the confinement.  Consequently, a discrete scaling symmetry only
exists for scattering lengths in the range $|\am|\ll |a|\ll l_z
$. Thus, for the moderate to strong confinements considered here,
where $l_z\sim |\am|$, the symmetry is clearly broken.

Figure~\ref{fig:3dstyle} shows a comparison between our q2D spectra
and the universal 3D results.  For increasing $|\am|/a$, we see that
the ground state trimer eventually resembles the 3D result even when
subjected to a strong confinement, $\cz =1$, which is consistent with
the approximately spherical shape of the trimer in
Fig.~\ref{fig:aspect}. This is reasonable since the properties of the
deepest trimer in 3D are set by short-distance physics: universal
theory predicts the size to be of order $|\am|$ for negative
scattering lengths $|\am|/a >-1$ \cite{Braaten2006}, and thus the
deepest trimer will only be weakly perturbed by the confinement when
$\cz \lesssim1$.  Another key feature of the q2D spectrum is the
raised threshold for free atom motion compared with the 3D case. As
expected, the trimers are significantly affected by the confinement
when their energies are well above the 3D threshold. However, note
that the binding energy of the ground state trimer can still be a
substantial fraction of $\omega_z$ when $|\am|/a \sim -1$, and thus
the trimer should be resistant to thermal dissociation when $T \ll
\omega_z$.

Remarkably, the raised q2D continuum threshold also stabilizes the two
deepest trimers for weak interactions, as clearly seen in
Fig.~\ref{fig:2dstyle}.  This results from the fact that the trimers
in 2D and 3D have the same $s$-wave symmetry and thus Efimov trimers
can smoothly evolve into long-range 2D-like trimers, without any level
crossings.  In the regime $|\am|/a < -1$, i.e., where no trimers exist
in 3D, we observe how a continuous scaling symmetry is recovered and
the trimer energies approach the universal 2D results.  For
sufficiently weak confinement, we even obtain avoided crossings, as
clearly observed in Fig.~\ref{fig:2dstyle} when $\cz =0.4$: here, a
third trimer appears for a scattering length close to $a_-$, a remnant
of the crossing of the deepest trimer with the continuum in 3D.  This
suggests that we can have a superposition of 2D and 3D-like trimers;
this is made possible by the presence of a repulsive barrier (see
Fig.~\ref{fig:pots}), as we discuss below. The third trimer state is
very weakly bound for $\cz=0.4$ and is expelled into the continuum as
the strength of the confinement is increased. Once $\cz\gtrsim 0.6$,
the third trimer disappears along with any pronounced avoided
crossings.

%Once $\cz\gtrsim 0.6$, the
%third trimer disappears along with any pronounced avoided crossings.

\section{Three-body potentials and trimer wavefunctions}
Considerable insight into the q2D spectra can be gained from the
adiabatic hyperspherical approach. This has been developed for both
3D~\cite{Nielsen2001,Braaten2006} and 2D~\cite{2Dreview}, and in
Appendix \ref{app:hs} we describe how this framework can be suitably
adapted to the intermediate q2D system. The hyperspherical approach
allows us to determine an effective three-body potential $V(R)$, where
the hyperradius $R^2=r_1^2+r_2^2+r_3^2$ is defined in terms of the
atom positions ${\bf r}_i$ at vanishing centre-of-mass coordinate.
The potential appears in an effective hyperradial Schr{\"o}dinger
equation $\left[-(1/2m)\partial^2/\partial R^2+V(R)\right]f_0(R)= (E_3
+\omega_z) \,f_0(R)$, where we ignore all but the lowest scattering
channel, an approximation valid in the 3D regime $R < l_z$, both when
$R \ll |a|$ and $R \gg |a|$.  For all other $R$, this should at least
provide a qualitative description --- in particular, we recover 2D
behaviour when $R \gg l_z$.  Note that $V(R)$ depends only on $l_z/a$
and makes no reference to the three-body parameter. Therefore, one
needs to supplement this with a short-distance boundary condition on
the wavefunction, which is equivalent to fixing $\Lambda$ or $|\am|$.

\begin{center}
\begin{figure*}[ht]
\vskip 0 pt
\includegraphics[width= 2\columnwidth, clip]{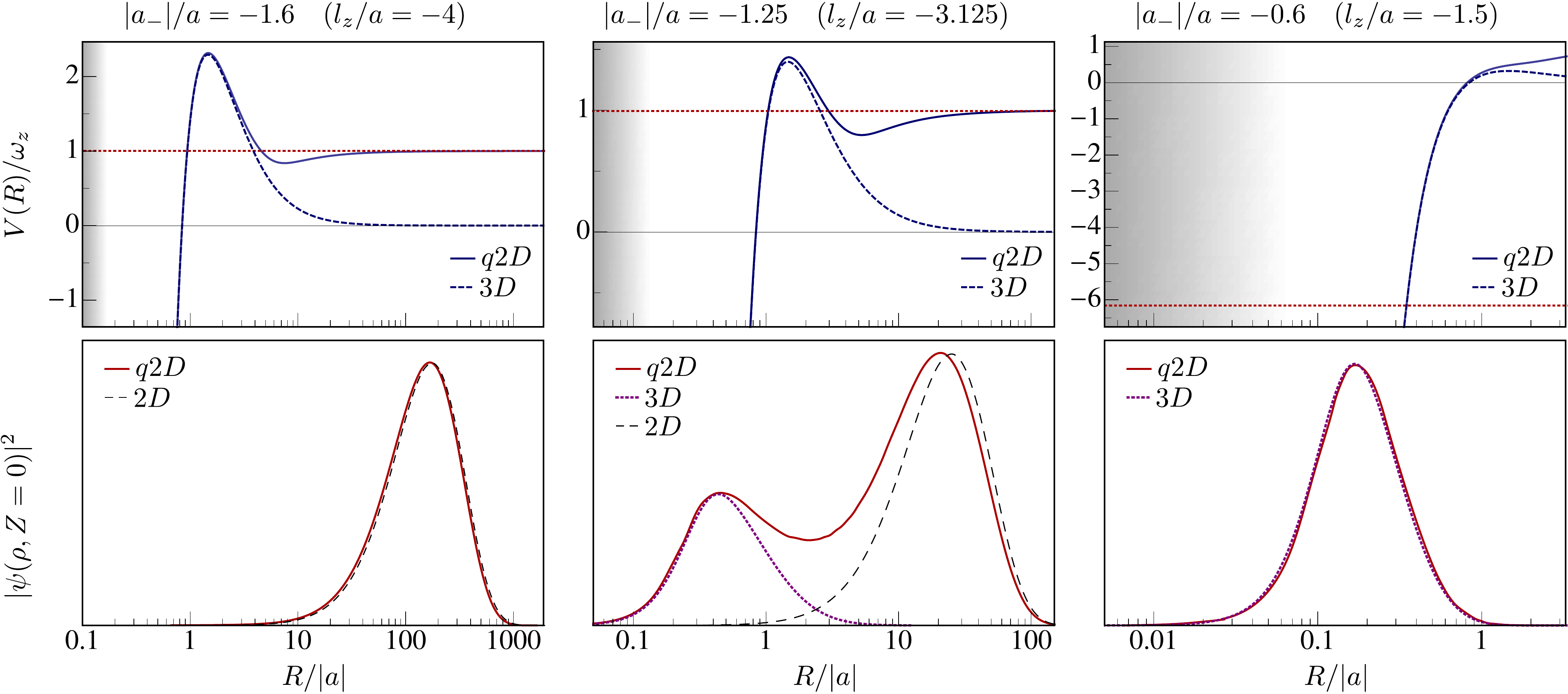}
\caption{Three-body potentials and corresponding wavefunctions of the
  deepest trimer.  The upper panels show the q2D adiabatic
  hyperspherical potential $V(R)$ for three different interactions
  $|\am|/a$, assuming $\cz =0.4$.  For reference, we also show the 3D
  potential. Note that the potentials match in the region $R < l_z$.
  The shaded area corresponds to the regime of short-distance physics
  where $R$ is less than the van der Waals range $\sim 1/\Lambda$.
  The resulting energies of the ground state trimer are shown as
  dashed horizontal lines. In the lower panels, the in-plane q2D
  probability densities $|\psi(\rho,Z=0)|^2$ are compared with the
  probability densities expected in purely 2D and 3D. This illustrates
  three qualitatively different interaction regimes: (left), for weak
  interactions, the trimer resides in the long-range attractive tail
  of the potential and matches the 2D result; (centre), for
  intermediate interactions, $a \sim \am$, the trimer configuration is
  a superposition of trimers of 2D and 3D character. Here, as there is
  no Efimov trimer at this scattering length in 3D, the 3D probability
  density is instead evaluated at $|\am|/a$ slightly larger than
  -1. (Right), for stronger interactions $|\am|/a >-1$ the trimer
  wavefunction resides mainly in the regime $R < |a|$ and closely
  resembles the 3D wavefunction.}
\label{fig:pots}
\end{figure*}
\end{center}

We show in Fig.~\ref{fig:pots} our calculated q2D hyperspherical
potentials for $|\am|/a <0$. At short distances, $V(R)$ matches the 3D
potential, which is attractive $\sim 1/R^2$ for $R\ll|a|$. On the
other hand, when $R \to \infty$, the effects of confinement become
apparent and $V(R) \to \omega_z - |E_b|$, corresponding to the free
motion of an atom and a dimer within the plane. When
$l_z/a\lesssim-2.5$, the potential also features a barrier at
intermediate radii $R\approx |a|$, with height $\sim
0.15/ma^2$. Approaching the limit $l_z/a \ll -1$, we see that $V(R)$
at $R\gg l_z$ resembles the 2D hyperspherical
potential~\cite{2Dreview}, which features a centrifugal repulsion and
a long-range attractive tail with respect to the continuum.  This
attraction gives rise to the two trimer states that exist for
arbitrarily weak interactions, unlike the Efimov trimers in 3D.  Note
that this differs from the three-body problem under \emph{isotropic}
harmonic
confinement~\cite{Jonsell2002,Werner2006,Thogersen2008,Portegies2011},
where we expect $V(R) \sim R^2$ for large $R$.

The presence of a repulsive barrier in the hyperspherical potential
means that trimer resonances at short distances can exist.  This same
feature is responsible for the loss resonances in 3D when $a \sim
\am$. For moderate q2D confinement (e.g., $\cz =0.4$ as shown in
Fig.~\ref{fig:pots}), it gives rise to a superposition of short-range
3D-like and long-range 2D-like trimer configurations. This is best
illustrated by considering the wavefunction $\psi({\boldsymbol
  \rho},Z)\equiv R^{3/2}\sum_{\k,N}e^{i\k\cdot
  {\boldsymbol\rho}}\phi_N(Z)\chi_\k^N$, describing the relative
motion of an atom with a pair. Here, $\boldsymbol \rho$ is the
atom-pair separation in the plane, $Z$ is the separation in the
transverse direction, and $\phi_N(Z)$ is the harmonic oscillator
wavefunction of a particle of mass $2m/3$, the reduced mass of the
atom-pair system. Note, in particular, that the probability
distribution $|\psi({\boldsymbol \rho},Z)|^2$ reduces to $|f_0(R)|^2$
in the 3D limit (up to normalisation factors), where $R^2 =
\frac{2}{3}(\rho^2+Z^2)$.

Figure~\ref{fig:pots} clearly illustrates that the wavefunction of the
deepest trimer exhibits both 2D and 3D-like components once $a$
approaches $\am$ and a trimer resonance in the 3D potential
appears. This hybridisation of trimer configurations is, in turn,
connected with the avoided crossings observed in
Fig.~\ref{fig:2dstyle}.  With increasing interaction, the deepest
trimer evolves into an Efimov-like trimer that resides at short
distances $R < |a|$.  This evolution from 2D to 3D behaviour is
mirrored by the aspect ratio $2\left<Z^2\right>/\left<\rho^2\right>$
(see Appendix \ref{app:AR}) in Fig.~\ref{fig:aspect}.  In general, we
expect to encounter a 2D-3D trimer hybridisation every time a 3D
trimer resonance appears behind the potential barrier. Thus, if we
were to relax the confinement until $\cz < 1/22.7$, the scenario
depicted in Fig.~\ref{fig:pots} would similarly occur for the second
deepest trimer.

\section{Consequences for Bose gas experiments in confined geometries}
We now discuss the ramifications of our results for current cold-atom
experiments. For definiteness, we consider $^{133}$Cs, the first
atomic species in which Efimov physics was
observed~\cite{Kraemer2006}. In this case, $a_-\simeq-957
a_0$~\cite{Berninger2011}, where $a_0$ is the Bohr radius, so that
$\cz=1$ corresponds to a confinement frequency of
$\omega_z\simeq2\pi\times30$kHz. While frequencies close to 100kHz
have been used for sideband cooling of
$^{133}$Cs~\cite{Vuletic1998,Bouchoule1999}, such strong confinement
is far from common in experiments investigating low-dimensional
physics. Indeed, scale invariance and universality in the repulsive 2D
Bose gas were investigated using a confinement
$\omega_z\simeq2\pi\times1.9$kHz, corresponding to $\cz \simeq
0.25$~\cite{Hung2011}.  Thus we expect that in realistic q2D
experiments, the energy of the deepest trimer will typically resemble
the universal 3D case.

On the other hand, the raised three-atom continuum under confinement
will strongly impact the 3D Efimov loss features in the three-atom
scattering. In particular, there can no longer be any trimer loss
resonances once $\omega_z$ exceeds the height of the repulsive barrier
in the hyperspherical potential --- we estimate this to occur when
$l_z/a\simeq-2.5$.  Thus, we expect the loss peak of the deepest
trimer to disappear for confinements
$\cz=|a_-|/l_z\gtrsim0.4$. Likewise, universality dictates that the
peak associated with the next Efimov state disappears when
$\cz\gtrsim0.4/22.7$. For the case of $^{133}$Cs, this latter value
corresponds to a confinement of $\omega_z\simeq2\pi\times9$Hz,
indicating that a very weak trapping potential is needed in order to
observe the second Efimov peak.  Note, further, that these arguments
translate in a generic manner to all geometries; one simply needs to
compare the height of the repulsive barrier at the 3D Efimov loss
resonance to the increase in the three-atom continuum.  Consequently,
our findings have implications for any experiment seeking to detect
shallow trimers in a trap, and hence the quest to observe true Efimov
scaling in an ultracold atomic gas.

To experimentally probe our q2D spectra in Figs.~\ref{fig:3dstyle} and
\ref{fig:2dstyle}, one could associate trimers using a radio frequency
pulse.  In 3D, this has successfully been applied by driving
transitions in the internal states of the atoms~\cite{Lompe2010} or by
modulating the magnetic field~\cite{Machtey2012}.  However, in
contrast to 3D, the resulting trimers in the strongly confined
geometry may be relatively long lived: hybrid trimers in the regime
$a\sim\am$ only have a small weight in the short distance region (see
Fig.~\ref{fig:pots}), resulting in a reduced overlap with deeply bound
(non-universal) states.  Accordingly, collisional relaxation into
deeper states (the main loss mechanism in 3D) will be suppressed.

\section{Conclusions and outlook}

A priori, there is no reason to believe that the physical picture
presented here should apply uniquely to three identical bosons in a
q2D geometry. For instance, we expect a trimer spectrum similar to
those of Figs.~\ref{fig:3dstyle} and~\ref{fig:2dstyle} to occur in a
q1D geometry --- theoretically this may be studied by including all
confinement levels instead of simply projecting onto the lowest level
as in Ref.~\cite{Mora2005}.  Likewise, while two tetramers have been
predicted~\cite{Stecher2008} and observed~\cite{Ferlaino2009} to
accompany each Efimov trimer in 3D, exactly two universal tetramers
are predicted to exist in 2D~\cite{Platter2004}. Consequently, we
expect a spectrum which displays avoided crossings between tetramers
of 2D and 3D character, while the two deepest tetramers are stabilized
by the application of a strong confinement. An important implication
of our work is thus that, under realistic experimental conditions,
three- and four-body correlations in the q1D and q2D Bose gas may be
strongly affected by Efimov physics, i.e., markedly different from
predictions of universal 1D and 2D theory.

Finally, our work suggests that strong q2D confinement could be used
to engineer more stable, Efimov-like, hybrid trimers owing to the
presence of a repulsive barrier and the associated small weight of the
trimer wavefunction at small distances.  In particular, our results
are also applicable to three distinguishable \emph{fermions} with
approximately equal interspecies interactions, as can be the case with
$^6$Li atoms~\cite{Lompe2010}.  This may allow for the formation of a
many-body state of long-lived trimers, an important goal which has so
far remained elusive in the context of ultracold experiments.

%%%%%%%%%%%%%%%%%%%%%%%%%%%%%%%%%%%%%%%%%%%%%%%%%
\acknowledgments

We gratefully acknowledge fruitful discussions with M.~Berninger,
E.~Braaten, F.~Ferlaino, R.~Grimm, B.~Huang, V.~Ngampruetikorn,
D.~S.~Petrov, M.~Zaccanti and A.~Zenesini. JL and PM wish to
acknowledge support from POLATOM. JL further acknowledges support from
the Carlsberg Foundation, PM from ERC AdG OSYRIS, EU EQuaM, and
Fundaci\'o Cellex, while MMP acknowledges support from the EPSRC under
Grant No.\ EP/H00369X/2.
 
%%%%%%%%%%%%%%%%%%%%%%%%%%%%%%%%%%%%%%%%%%%%%%%%%

\appendix
\section{\label{app:H}Hamiltonian and two-body problem}
The Hamiltonian in the absence of confinement is chosen to be
\begin{align}
{\cal H} = \sum_{\k}\ek\hat a^\dagger_{\k}\hat a_{\k}
+\frac{g}{2}
\sum_{{\k_1,\k_2,\k_3}}
\xi(k_{12}) \xi(k_{34})
\hat a^\dag_{\k_1} \hat a^\dag_{\k_2}
\hat a_{\k_3} \hat a_{\k_4},
\nn
\end{align}
where $\hat a_\k$ ($\hat a^\dagger_\k$) is the annihilation (creation)
operator of atoms with momentum $\k$, $\k_{ij}\equiv
\frac{\k_i-\k_j}2$ is the relative momentum, and
$\k_4=\k_1+\k_2-\k_3$. $\xi(k)$ is a function describing the cutoff of
the interaction at large momenta, and we take this to be
$\xi(k)=e^{-k^2/\Lambda^2}$.

The two-body T-matrix appearing in the STM equation describes
scattering of two atoms with total planar momentum $\k$ and energy $E$
measured from the two-atom continuum. It takes the form
\begin{align*}
  \T(\k,E)= &\frac{2\sqrt{2\pi}}{m}\left\{\frac{l_z}{a} -\F
    \left(\frac{-E+k^2/4m}{\op}\right)\right\}^{-1}.
\label{eq:Tq2d}
\end{align*}
Here, the interaction is renormalized by the use of the 3D scattering
length $a$. For a Gaussian cutoff, we obtain
\begin{align*}
\F(x)=&\int_0^\infty 
\mbox{d}u
\frac{1-\frac{e^{-xu}}
{\sqrt{\left[(1+\lambda)^2-(1-\lambda)^2e^{-2u}\right]/(2u+4\lambda)}}}
{\sqrt{4\pi(u+2\lambda)^3}},
\end{align*}
with $\lambda\equiv (l_z\Lambda)^{-2}$. Our expression for $\F$
reduces to that of Ref.~\cite{review2008} in the limit $\lambda\to0$.
The dimer binding energy $E_b$ measured from the continuum is defined
through $l_z/a=\F(-E_b/\omega_z)$.

In our model, the wavefunctions of the relative motion evaluated at
the origin are
\begin{align*}
f_{n_r} & =(-1)^{n_r/2}\sqrt{\frac{(n_r-1)!!}{n_r!!}} \frac{1}
{\sqrt{1+\lambda}} \left(\frac{1-\lambda}{1+\lambda}\right)^{n_r/2},
\end{align*}
if $n_r$ is even, and 0 otherwise (we absorbed the prefactor
$(m\omega_z/2\pi)^{1/4}$ into the definition of $\T$).

\section{\label{app:3bd}Three-body problem}
In Eq.~(\ref{eq:stmq2d}) we employ the simplification
$\xi(\k_1+\k_2/2)\xi(\k_2+\k_1/2)\rightarrow \xi(k_1)\xi(k_2)$ in the
2D plane. This allows us to project the equation analytically onto the
$s$-wave, $\chi_{k_1}^{N_1}\equiv \int
\frac{d\phi}{2\pi}\chi_{\k_1}^{N_1}$, where $\phi$ is the angle of
$\k_1$ with some axis.  Once $\Lambda$ has been used to fix the
three-body parameter, the physics at energy scales much smaller than
$\Lambda^2/m$ becomes insensitive to this change. Using the 3D STM
equation with our two-body interaction, for the deepest Efimov trimer
we find the energy $-0.05\Lambda^2/m$ at the Feshbach resonance and
$\am=-9.39/\Lambda$. Since the trimer energies considered in the q2D
geometry are always smaller than those in the 3D geometry (see
Fig.~\ref{fig:3dstyle}), the assumption that we consider energies much
smaller than $\Lambda^2/m$ is well justified.

\section{\label{app:hs}Hyperspherical approach in q2D}
Beginning with the Jacobi coordinates $\vect{r}_{ij} = \vect{r}_i -
\vect{r}_{j}$ and $\vect{r}_{k,ij} = \frac{1}{2} (\vect{r}_i +
\vect{r}_j)-\vect{r}_k$, and assuming that $\vect{r}_1 + \vect{r}_2 +
\vect{r}_3 = 0$, the hyperradius corresponds to $R^2 = \frac{1}{2}
r^2_{ij} + \frac{2}{3} r^2_{k,ij}$, while the transformation $r_{ij} =
\sqrt{2} R \sin\alpha_k$, \ $r_{k,ij} = R \sqrt{\frac{3}{2}}
\cos\alpha_k$ defines the hyperangle $\alpha_k$.  Following
Ref.~\cite{Braaten2006}, we use the hyperspherical expansion of the
wave function: $\Psi(R,\Omega) = \frac{1}{R^{5/2}\sin(2\alpha_k)}
\sum_{n=0}^\infty f_n(R) \Phi_n(R,\Omega)$.  Here, the angular
quantity $\Omega = (\alpha_k,\theta_{ij},\theta_{k,ij})$, where the
projections $\hat{\vect{r}}_{ij}\cdot\hat{z} = \cos\theta_{ij}$, \
$\hat{\vect{r}}_{k,ij}\cdot\hat{z} = \cos\theta_{k,ij}$, and we assume
that $\Psi$ is approximately independent of the azimuthal angles.  We
now further expand the angular function $\Phi_n(R,\Omega) =
\sum_{\vect{m}} \eta_{n\vect{m}}(R,\alpha_k) \ h_\vect{m}(R,\Omega)$,
with $\vect{m} = (m_1,m_2)$ a set of non-negative integers. Writing
$h_{\vect{m}} = \tau_{m_1}(R\sin\alpha_k,\theta_{ij}) \times
\tau_{m_2}(R\cos\alpha_k,\theta_{k,ij})$, the function
$\tau(X,\theta)$ obeys the equation
\begin{align*}
  \lbc - \frac{l_z^2}{X^2} \frac{1}{\sin\theta}
  \frac{\partial}{\partial\theta} \lba \sin\theta
  \frac{\partial}{\partial\theta} \rba + \frac{X^2}{l_z^2}
  \cos^2\theta \rbc \tau = 2 \mu \tau
\end{align*}
where $\mu$ is an eigenvalue that is independent of $\theta$, and $X =
R\cos\alpha_k$ or $X=R\sin\alpha_k$, depending on whether we consider
$\theta_{k,ij}$ or $\theta_{ij}$. In the limit $X \ll l_z$, $\tau$
yields the Legendre polynomials expected in the 3D case, while in the
opposite limit $X \gg l_z$, $\tau$ evolves into the harmonic
oscillator wavefunctions of the q2D confinement. Finally we use $\mu$
to solve the equation for $\eta_{0\vect{0}}$ and obtain the lowest
hyperspherical potential $V(R)$ in the usual way within the adiabatic
hyperspherical approximation~\cite{Braaten2006}.

\section{\label{app:AR}Aspect ratio of trimers}
The wavefunction can in general be decomposed in its Fadeev components
$ \Psi(\vect{r}_1,\vect{r}_2,\vect{r}_3)=
\psi^{(1)}(\vect{r}_{23},\vect{r}_{1,23})+
\psi^{(2)}(\vect{r}_{13},\vect{r}_{2,31})+
\psi^{(3)}(\vect{r}_{12},\vect{r}_{3,12})$. Here
$\psi^{(1)}(\vect{r}_{23},\vect{r}_{1,23})$ is the real space form of
the function $\psi_{\k_{23},\k_1,n_{23},N_1}^{(1)}\propto
\frac{f_{n_{23}}\chi_{\k_1}^{N_1}}{E_3-k_{23}^2/m-3k_1^2/4m-(n_{23}+N_1)\op}$.
We then calculate the aspect ratio of the relative atom-pair
coordinate, assuming that cross terms may be neglected:
\begin{align}
  \frac{2\langle z_{1,23}^2\rangle}{\langle \rho_{1,23}^2\rangle} & =
  \frac{2\int
    d^2\rho_{23}d^2\rho_{1,23}dz_{23}dz_{1,23}z_{1,23}^2|\psi^{(1)}(\vect{r}_{23},\vect{r}_{1,23})|^2}
  {\int
    d^2\rho_{12}d^2\rho_{1,23}dz_{23}dz_{1,23}\rho^2_{1,23}|\psi^{(1)}(\vect{r}_{23},\vect{r}_{1,23})|^2}.
  \nn
\end{align}
When $|\am|/a\ll-1$, the aspect ratio approaches the 2D limit where
$\langle z_{1,23}^2\rangle=3 l_z^2/4$ and $\langle \rho_{1,23}^2\rangle
=0.037/(m|E_b|)$.

\bibliography{quasi2defimov}%,revtex-custm}

\begin{thebibliography}{10}

\bibitem{Efimov1970}
V. Efimov, Phys. Lett. B {\bf 33},  563  (1970).

\bibitem{Efimov1971}
V. Efimov, Sov. J. Nucl. Phys. {\bf 12},  589  (1971).

\bibitem{Ferlaino2011}
F. Ferlaino {\it et~al.}, Few-Body Systems {\bf 51},  113  (2011).

\bibitem{Nishida2013}
Y. Nishida, Y. Kato, and C.~D. Batista, Nature Phys. {\bf 9},  93  (2013).

\bibitem{Kraemer2006}
T. {Kraemer} {\it et~al.}, Nature {\bf 440},  315  (2006).

\bibitem{Braaten2006}
E. Braaten and H.-W. Hammer, Physics Reports {\bf 428},  259  (2006).

\bibitem{mandelbrot1983fractal}
B. Mandelbrot, {\em The Fractal Geometry of Nature} (Henry Holt and Company,
  New York, 1983).

\bibitem{Hadzibabic2006}
Z. {Hadzibabic} {\it et~al.}, Nature {\bf 441},  1118  (2006).

\bibitem{Clade2009}
P. Clad\'e {\it et~al.}, Phys. Rev. Lett. {\bf 102},  170401  (2009).

\bibitem{Hung2011}
C.-L. {Hung}, X. {Zhang}, N. {Gemelke}, and C. {Chin}, Nature {\bf 470},  236
  (2011).

\bibitem{Nishida2011}
Y. Nishida and S. Tan, Few-Body Systems {\bf 51},  191  (2011).

\bibitem{Bruch1979}
L.~W. Bruch and J.~A. Tjon, \pra {\bf 19},  425  (1979).

\bibitem{Sommer2012}
A.~T. Sommer {\it et~al.}, Phys. Rev. Lett. {\bf 108},  045302  (2012).

\bibitem{Guenter2005}
K. G\"unter {\it et~al.}, Phys. Rev. Lett. {\bf 95},  230401  (2005).

\bibitem{Petrov2001}
D.~S. Petrov and G.~V. Shlyapnikov, Phys. Rev. A {\bf 64},  012706  (2001).

\bibitem{zaccanti2009}
M. Zaccanti {\it et~al.}, Nature Phys. {\bf 5},  586  (2009).

\bibitem{Gross2009}
N. Gross, Z. Shotan, S. Kokkelmans, and L. Khaykovich, Phys. Rev. Lett. {\bf
  103},  163202  (2009).

\bibitem{Pollack2009}
S.~E. Pollack, D. Dries, and R.~G. Hulet, Science {\bf 326},  1683  (2009).

\bibitem{Berninger2011}
M. Berninger {\it et~al.}, Phys. Rev. Lett. {\bf 107},  120401  (2011).

\bibitem{Roy2013}
S. Roy {\it et~al.}, Phys. Rev. Lett. {\bf 111},  053202  (2013).

\bibitem{Wang2012}
J. Wang, J.~P. D'Incao, B.~D. Esry, and C.~H. Greene, Phys. Rev. Lett. {\bf
  108},  263001  (2012).

\bibitem{stm}
G.~V. Skorniakov and K.~A. Ter-Martirosian, Sov. Phys. JETP {\bf 4},  648
  (1957).

\bibitem{Levinsen2009}
J. Levinsen, T.~G. Tiecke, J.~T.~M. Walraven, and D.~S. Petrov, Phys. Rev.
  Lett. {\bf 103},  153202  (2009).

\bibitem{Levinsen2013}
J. Levinsen and M.~M. Parish, Phys. Rev. Lett. {\bf 110},  055304  (2013).

\bibitem{Wigner}
E.~P. Wigner, {\em Group Theory and Its Application to the Quantum Mechanics of
  Atomic Spectra} (Academic Press, New York, 1959).

\bibitem{SchwingerQM}
J. Schwinger,  in {\em Quantum Theory of Angular Momentum}, edited by L.~C.
  Biedenharn and H. {van Dam} (Academic Press, New York, 1965), pp.\ 229--279.

\bibitem{Nielsen2001}
E. {Nielsen}, D.~V. {Fedorov}, A.~S. {Jensen}, and E. {Garrido}, Physics
  Reports {\bf 347},  373  (2001).

\bibitem{2Dreview}
E. Nielsen, D.~V. Fedorov, and A.~S. Jensen, Few-Body Systems {\bf 27},  15
  (1999).

\bibitem{Jonsell2002}
S. Jonsell, H. Heiselberg, and C.~J. Pethick, Phys. Rev. Lett. {\bf 89},
  250401  (2002).

\bibitem{Werner2006}
F. Werner and Y. Castin, Phys. Rev. Lett. {\bf 97},  150401  (2006).

\bibitem{Thogersen2008}
M. Th\o{}gersen, D.~V. Fedorov, and A.~S. Jensen, Phys. Rev. A {\bf 78},
  020501  (2008).

\bibitem{Portegies2011}
J. Portegies and S. Kokkelmans, Few-Body Systems {\bf 51},  219  (2011).

\bibitem{Vuletic1998}
V. Vuleti\ifmmode~\acute{c}\else \'{c}\fi{}, A.~J. Kerman, C. Chin, and S. Chu,
  Phys. Rev. Lett. {\bf 82},  1406  (1999).

\bibitem{Bouchoule1999}
I. Bouchoule {\it et~al.}, Phys. Rev. A {\bf 59},  R8  (1999).

\bibitem{Lompe2010}
T. Lompe {\it et~al.}, Science {\bf 330},  940  (2010).

\bibitem{Machtey2012}
O. Machtey, Z. Shotan, N. Gross, and L. Khaykovich, Phys. Rev. Lett. {\bf 108},
   210406  (2012).

\bibitem{Mora2005}
C. Mora, R. Egger, and A.~O. Gogolin, Phys. Rev. A {\bf 71},  052705  (2005).

\bibitem{Stecher2008}
J. {von Stecher}, J.~P. {D'Incao}, and C.~H. {Greene}, Nature Physics {\bf 5},
  417  (2009).

\bibitem{Ferlaino2009}
F. Ferlaino {\it et~al.}, Phys. Rev. Lett. {\bf 102},  140401  (2009).

\bibitem{Platter2004}
L. Platter, H.-W. Hammer, and U.-G. Mei{\ss}ner, Few-Body Systems {\bf 35},
  169  (2004).

\bibitem{review2008}
I. Bloch, J. Dalibard, and W. Zwerger, Rev. Mod. Phys. {\bf 80},  885  (2008).

\end{thebibliography}

\end{document}